\documentclass[aps,pra,reprint,10pt,superscriptaddress,amssymb,amsfonts,longbibliography,nofootinbib]{revtex4-1}

\usepackage{amsmath}
\usepackage{graphicx}
\usepackage{dcolumn}
\usepackage{bbm}
\usepackage{subfigure}
\usepackage{amssymb}
\usepackage{wasysym}
\usepackage{MnSymbol}
\usepackage{color}
\usepackage{mathrsfs}
\usepackage[normalem]{ulem}

\newcommand{\beq}{\begin{equation}}
\newcommand{\eeq}{\end{equation}}

\stepcounter{secnumdepth}
\stepcounter{tocdepth}

\begin{document}

\title{Pinch Point Singularities of Tensor Spin Liquids}

\author{Abhinav Prem}
\affiliation{Department of Physics and Center for Theory of Quantum Matter, University of Colorado, Boulder, CO 80309, USA}

\author{Sagar Vijay}
\affiliation{Department of Physics, Massachusetts Institute of Technology, Cambridge, MA 02139, USA}

\author{Yang-Zhi Chou}
\affiliation{Department of Physics and Center for Theory of Quantum Matter, University of Colorado, Boulder, CO 80309, USA}

\author{Michael Pretko}
\affiliation{Department of Physics and Center for Theory of Quantum Matter, University of Colorado, Boulder, CO 80309, USA}

\author{Rahul M. Nandkishore}
\affiliation{Department of Physics and Center for Theory of Quantum Matter, University of Colorado, Boulder, CO 80309, USA}

%\date{\today}
\begin{abstract}
Recently, a new class of three-dimensional spin liquid models have been theoretically discovered, which feature generalized Coulomb phases of emergent symmetric tensor $U(1)$ gauge theories. These ``higher rank'' tensor models are particularly intriguing due to the presence of quasi-particles with restricted mobility, such as fractons. We investigate universal experimental signatures of tensor Coulomb phases. Most notably, we show that tensor Coulomb spin liquids (both quantum and classical) feature characteristic pinch-point singularities in their spin-spin correlation functions, accessible via neutron scattering, which can be readily distinguished from pinch points in conventional $U(1)$ spin liquids. These pinch points can thus serve as a crisp experimental diagnostic for such phases. We also tabulate the low-temperature heat capacity of various tensor Coulomb phases, which serves as a useful additional diagnostic in certain cases.
\end{abstract}

\maketitle

\section{Introduction}

Quantum spin liquids describe exotic, interacting spin systems, in which quantum fluctuations prevent conventional magnetic ordering all the way down to zero temperature. These phases are characterized by a pattern of long-range quantum entanglement in their ground states and the presence of exotic fractionalized excitations. Spin liquids are believed to occur in gapped and gapless varieties~\cite{gapless1,gapless2,kagome,banerjee1,balz,naturebalents}, and are theoretically well-described as emergent gauge theories~\cite{lucile,review2}.

The gauge theory description of a spin liquid can take a number of different forms, ranging from intricate string-net models~\cite{stringnet} to familiar $U(1)$ Maxwell theory. The latter case has a number of promising experimental candidates in the form of the ``spin ice" pyrochlore materials, including the classical spin ices Dy$_2$Ti$_2$O$_7$ and Ho$_2$Ti$_2$O$_7$, as well as the quantum spin ices Yb$_2$Ti$_2$O$_7$ and Pr$_2$Zr$_2$O$_7$~\cite{pyrochlore1,pyrochlore2,lucile,spinice,qspinice,pr1,pr2,pr3}. Over a range of low temperatures, these materials exist in a symmetry-preserving phase consistent with the expected behavior of a deconfined Coulomb phase of an emergent $U(1)$ gauge field. It is possible that in some materials, this emergent electromagnetism may survive down to zero temperature, providing an example of a $U(1)$ quantum spin liquid. Regardless, we may conclusively identify these materials as having at least classical spin liquid behavior: resisting symmetry breaking down to unusually low temperatures due to frustration between many energetically equivalent classical configurations.

Conventional $U(1)$ spin liquids exhibits striking experimental signatures. Most notably, the Coulomb phase of a $U(1)$ gauge theory exhibits characteristic ``pinch-point" singularities in its correlation functions; these pinch-points may be observed in spin-spin correlation functions that are readily measured via neutron scattering experiments~\cite{henley1,henley2,pinch}. In the quantum spin liquid, these singularities arise as a direct consequence of the gapless excitations of the system, corresponding to the emergent photon of the $U(1)$ gauge theory. Such singularities are absent in gapped quantum spin liquids.

While a conventional $U(1)$ spin liquid, described in terms of an emergent Maxwell theory, has been a subject of intense theoretical and experimental study during the past two decades~\cite{lucile,review2,pyrochlore1,pyrochlore2,henley1,henley2,pinch,motrunich,
moessner,hermele,stories,ether,banerjee,ross,savary1,savary2,chong,savary3,me,
perspective,savary4,liujun1,liujun2}, recent theoretical developments have uncovered a new class of generalized $U(1)$ spin liquids which we have only just begun to understand~\cite{cenke,alex,sub,genem,mach,screening,theta}. Instead of the familiar vector $U(1)$ gauge field of Maxwell theory, these three-dimensional spin systems are described by the deconfined ``Coulomb" phase of emergent symmetric \emph{tensor} $U(1)$ gauge fields\footnote{\emph{Anti}-symmetric $U(1)$ tensor gauge fields tend to not have deconfined phases in three or fewer dimensions~\cite{orland,pearson,rey,gerbe,lake}}.

This new class of spin liquids has some properties in common with the conventional $U(1)$ spin liquid, such as protected gapless gauge modes. What sets these new tensor gauge theories apart, however, is the behavior of the emergent, gapped charge excitations, which have severe restrictions on their mobility. The gauge charges can be restricted to motion within one- or two-dimensional subspaces, or in certain models, can be restricted from moving at all. These immobile, charged excitations (termed ``fractons''), as well as the  gapped excitations with reduced mobility were first obtained in completely \textit{gapped} three-dimensional systems with intricate patterns of long-ranged entanglement, and have since been encountered in a wide variety of physical systems~\cite{review,chamon,bravyi,haah,cast,yoshida,haah2,fracton1,fracton2,matter,glassy,
williamson,sagarlayer,hanlayer,parton,slagle,bowen,nonabel,balents,field,valbert,
correlation,simple,entanglement,bernevig,albert,generic1,generic2,z3,elasticity,
gromov,pai,higgs1,higgs2,subsystem,deconfined,foliated,ungauging,fractalsym,twisted,symfrac,uhaah}. {Gapped} fracton phases, such as Haah's code or the X-Cube model, display glassy quantum dynamics in their approach to equilibrium, which may serve as a useful diagnostic of these systems~\cite{glassy}. On the other hand, the ``generalized" Hall conductivity predicted for the two-dimensional chiral gapped tensor gauge theory~\cite{theta,matter} has been conjectured to be a manifestation of the torsional Hall viscosity~\cite{gromov}, providing another experimental signature of fracton physics.

While the gapless tensor $U(1)$ spin liquids have likewise been a topic of intense recent theoretical study, little is known about sharp experimental signatures of these systems.  In the present work, we will identify certain key signatures which can be used to diagnose the presence of different types of emergent tensor $U(1)$ spin liquids in experiments, placing particular emphasis on the rank-2 spin liquids, i.e. where the emergent gauge field in the system of interest is a two-component symmetric tensor. Such experimental metrics may provide important clues which guide the search for physical systems realizing tensor Coulomb behavior.

\begin{figure}[t!]
 \centering
 \includegraphics[width=0.5\textwidth]{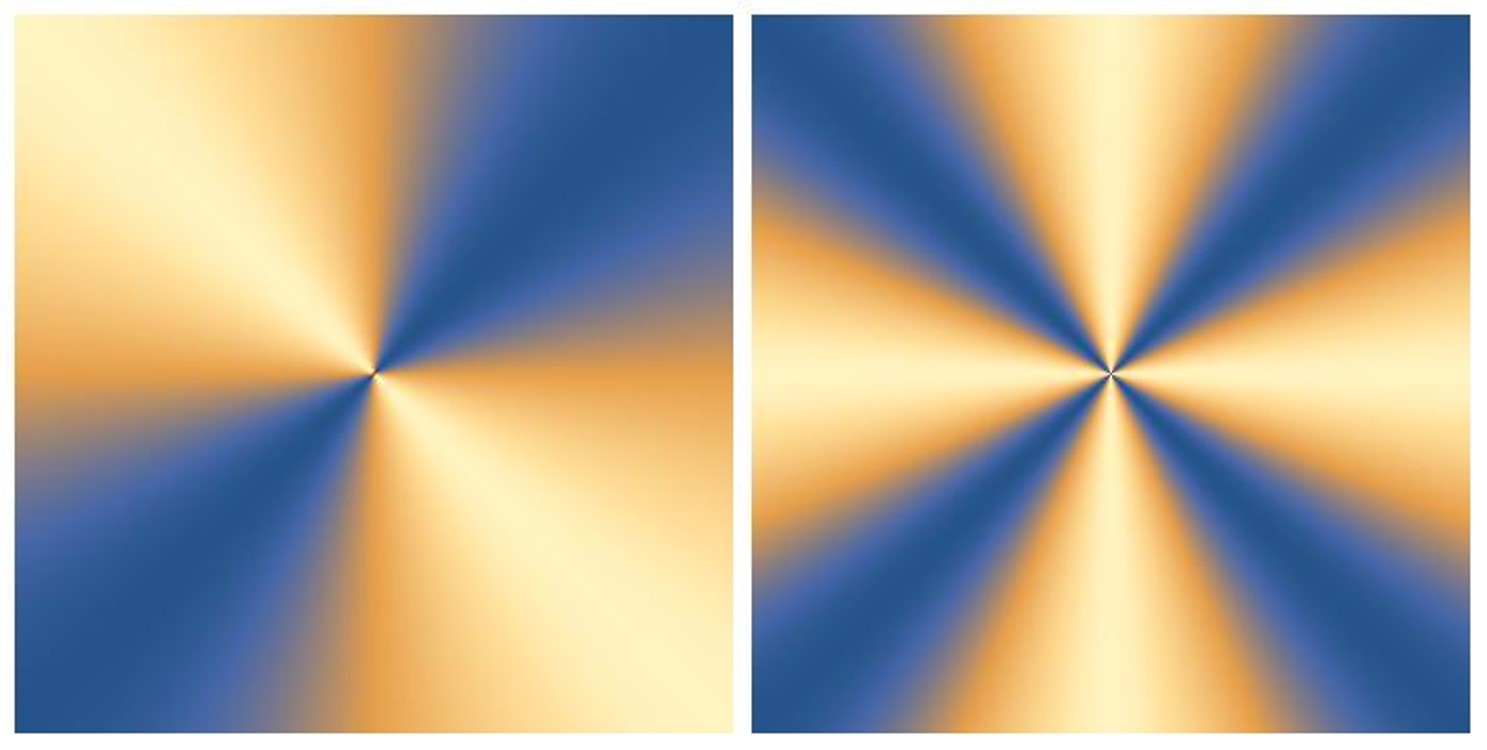}
 \caption{The pinch point singularities of a conventional $U(1)$ spin liquid (left) have a characteristic two-fold symmetry.  In contrast, pinch points of the rank-2 tensor spin liquids (right) have a characteristic four-fold symmetry, which should allow for easy distinction in neutron-scattering data.  (The two plots display $\langle E^x(q)E^y(-q)\rangle$ (left) and $\langle E^{xx}(q)E^{yy}(-q)\rangle$ (right), with cross-sections taken in the $q_z=0$ plane.)}
 \label{fig:pinch}
\end{figure}

Most notably, we study the behavior of the spin-spin correlation functions of these new spin liquids.  We begin by studying the ground state correlation functions of the quantum version of these spin liquids by making use of their low-energy effective field theory~\cite{alex,sub,genem}. We show that the spin-spin correlation functions exhibit a new pinch-point singularity due to the tensor nature of the gapless gauge excitations.  Certain features of these singularities are universal, independent of details at the lattice-scale, and are applicable to any microscopic model featuring a tensor Coulomb phase.  These universal features allow the tensor $U(1)$ spin liquids to be easily distinguished from a conventional $U(1)$ spin liquid in experiment. For rank-2 tensor models, we show that the pinch points have a characteristic \emph{four-fold} symmetry, as opposed to the two-fold symmetry of pinch points in more conventional spin liquids (see Figure~\ref{fig:pinch}), which will allow for straightforward detection in experiments. We note that the singularities generically remain pointlike in these systems, in contrast with the ``pinch-line" singularities seen in certain non-generic tensor spin liquid models~\cite{line}.  We then move on to study the finite-temperature spin-spin correlation functions of classical tensor Coulomb spin liquids, which we show have similar pinch-point singularities.

In addition to pinch-point singularities, we also tabulate the heat capacities of the various tensor $U(1)$ spin liquids, which should be readily accessible to experiments. The gapless modes of these models lead to a power-law contribution to the heat capacity, distinguishable from that of vector $U(1)$ models in certain cases. In a conventional $U(1)$ spin liquid, the linear dispersion of the gauge modes means that their contribution to the heat capacity cannot be easily separated from that of phonons, and hence does not serve as a useful experimental probe.  In contrast, some of the tensor $U(1)$ spin liquids have non-linearly dispersing gauge modes which provide a dominant contribution to the low-temperature heat capacity and should thus serves as a useful diagnostic for the tensorial nature of the gauge field. 

\section{Review of Tensor Gauge Theory}

We here review the basic properties of the two simplest tensor $U(1)$ quantum spin liquids and refer the reader to previous literature~\cite{alex,sub,genem,theta} for more details.  These spin liquids are described in terms of an emergent, symmetric rank-2 tensor field $A_{ij}$, where all indices refer to spatial coordinates $i,j=1,2,3$. This gauge field possesses a canonical conjugate variable $E_{ij}$, which corresponds to a generalized electric field. Multiple  theories may be defined using these variables, each of which is uniquely determined by the form chosen for Gauss's law. 

\subsection{Scalar Charge Theory}

One simple theory we can write down has a Gauss's law of the form
\begin{equation}
\partial_i\partial_j E^{ij} = \rho,
\end{equation}
for a scalar-valued charge $\rho$. Within the low-energy sector, the corresponding gauge transformation is
\begin{equation}
A_{ij}\rightarrow A_{ij}+\partial_i\partial_j\alpha,
\end{equation}
where $\alpha$ is a scalar function with arbitrary spatial dependence. This system admits gapless gauge modes, with a low-energy Hamiltonian given by
\begin{equation}
H = \int d^3x\frac{1}{2}(E^{ij}E_{ij} + B^{ij}B_{ij}),
\label{hamiltonian}
\end{equation}
where $B^{ij} = \epsilon^{iab}\partial_aA_b^{\,\,\,j}$ is the gauge-invariant magnetic field operator. This Hamiltonian leads to five gapless gauge modes, each with a linear dispersion $\omega\propto q$.

Much more notable than the gauge mode, however, is the charge sector of the theory, which has properties with no analogue in more conventional gauge theories. In addition to the conservation of charge
\begin{equation}
\int d^3x\,\rho = \textrm{constant},
\end{equation}
this theory also exhibits conservation of dipole moment
\begin{equation}
\int d^3x\,(\rho x^i) = \textrm{constant}.
\end{equation}
This conservation law has the severe immediate consequence that the fundamental charges of the theory are strictly immobile $i.e.$ are fracton excitations. Only charge-neutral bound states, such as dipolar bound states, are free to move around the system.

\subsection{Vector Charge Theory}

It is also possible to consider a slightly modified theory of a rank-2 tensor, with a different version of Gauss's law, which takes the form
\begin{equation}
\partial_iE^{ij} = \rho^j,
\end{equation}
for a vector-valued charge density $\rho^j$.  The corresponding low-energy gauge transformation is
\begin{equation}
A_{ij}\rightarrow A_{ij} + \partial_i\alpha_j + \partial_j\alpha_i,
\end{equation}
where $\alpha_i$ is a function with arbitrary spatial dependence. The low-energy Hamiltonian for the gauge sector of this theory takes the same form as in Eq.~\eqref{hamiltonian}, but with a modified magnetic field operator, $B^{ij} = \epsilon^{iab}\epsilon^{jcd}\partial_a\partial_c A_{bd}$. In this case, the Hamiltonian leads to three gapless quadratically dispersing gauge modes, $\omega\propto q^2$. This theory also possesses an unusual set of charge conservation laws, in that the vector charges obey not only conservation of charge,
\begin{equation}
\int d^3x\,\rho^i = \textrm{constant},
\end{equation}
but also a second conservation law pertaining to the angular moment of charge,
\begin{equation}
\int d^3x\,\epsilon^{ijk}\rho_jx_k = \textrm{constant}.
\end{equation}
This second conservation law has the unusual consequence that the vector charges are restricted to motion only in the direction of their charge vector, while motion in the perpendicular directions is ruled out by gauge invariance.  This causes the charges to behave like one-dimensional particles, despite being embedded in three-dimensional space.

\section{Pinch-Point Singularities}

In a spin liquid setting, the physical spin operators can be mapped directly onto gauge-invariant field operators of an emergent gauge theory. We here focus on the case where the spins map onto electric field operators (similar arguments apply when the spins are mapped onto magnetic operators.). We calculate the correlation function of a rank-2 electric tensor as
\begin{equation}
\langle E_{ij}(x)E_{k\ell}(0)\rangle
\end{equation}
and similarly for tensors of higher rank. The physical spin correlators in the long distance limit will be dominated by the correlation functions of the gapless gauge field, and given by some linear combination of these tensor correlation functions
\begin{equation}
\langle S_z(x) S_z(0)\rangle = \sum_{ijk\ell} C^{ijk\ell}\langle E_{ij}(x)E_{k\ell}(0)\rangle.
\label{super}
\end{equation}
Here the separation $x$ is implicitly large, and the structure factor $C^{ijk\ell}$ is constrained by the symmetries of the underlying lattice. In the following, we focus on the universal (long distance, small wavevector) behavior of these correlation functions, which should not depend on the precise form of the structure factors.

\subsection{Conventional $U(1)$ Spin Liquid}

For convenience, we recall the calculation of pinch-point singularities in a conventional $U(1)$ spin liquid, which will generalize naturally to the tensor case. The appropriate low-energy Hamiltonian takes the usual Maxwell form,
\begin{equation}
H = \int d^3x \frac{1}{2}(E^iE_i + B^iB_i),
\end{equation}
where $B^i = \epsilon^{ijk}\partial_jA_k$, and we implicitly have the gauge constraint $\partial_i E^i = 0$. In momentum space, the Hamiltonian decouples into independent harmonic oscillator modes, and thus by equipartition, the two terms of the Hamiltonian contribute equally to the ground state energy. For Maxwell theory, the dispersion is linear $\omega \propto q$, leading to a zero point energy proportional to $q$ for each mode, from which we conclude that
\begin{equation}
\langle E^i(q)E_i(-q)\rangle \propto q.
\end{equation}
We must now restore the full tensor structure. To do this, we start with the isotropic result $\delta^{ij}$, as if all modes were present, and then project out the divergence mode, which is absent from the low-energy sector. The final electric field correlator then takes the form
\begin{equation}
\label{vectorcorr}
\langle E^i(q)E^j(-q)\rangle \propto q\left(\delta^{ij}-\frac{q^iq^j}{q^2}\right). 
\end{equation}
The second term, arising due to the projection into the gauge sector, leads to ``pinch-point" singularities in the correlation function, in that the ratio $q^i q^j/q^2$ has different limits upon approaching the origin $\mathbf{q}=0$ from different directions, as depicted schematically in Figure~\ref{fig:pinch}. These singularities can be easily detected via neutron scattering, thereby serving as a powerful tool for diagnosing $U(1)$ spin liquids in experiments. We note that an important feature of the correlation function Eq.~\eqref{vectorcorr} is its two-fold symmetry, illustrated explicitly in e.g. the $\langle E^x(q)E^y(-q)\rangle$ correlator shown in Fig.~\ref{fig:pinch}.

One can also consider the finite-temperature behavior of correlation functions in a classical $U(1)$ spin liquid, where the quantum splitting of degeneracies is unimportant.  In this case, we impose the spin-ice constraint, $\partial_iE^i = 0$, but regard all states within this spin-ice manifold as being roughly energetically equivalent.  The free energy of the system is then set almost entirely by entropic effects, $F\approx -TS$.  By the central limit theorem, the probability distribution for the emergent electric field $E^i$ must be Gaussian in the thermodynamic limit~\cite{henley2}, such that
\begin{equation}
F/T = K \int d^3x\, E^iE_i, \label{modelice}
\end{equation}
for some constant $K$, where the spin-ice constraint $\partial_iE^i = 0$ is left implicit.  Were it not for this constraint, we could simply conclude that $\langle E_i(q)E_j(-q)\rangle\propto (1/K) \delta_{ij}$.  After projecting out the longitudinal component, however, the correct correlation function behaves as
\begin{equation}
\langle E_i(q)E_j(-q)\rangle_c \propto \frac{1}{K}\bigg(\delta_{ij} - \frac{q_iq_j}{q^2}\bigg),
\end{equation}
where the subscript $c$ denotes ``classical."  Note that this classical correlation function has the same singular tensor structure as the quantum case, but with a different pre-factor. Note also that the pinch point structure comes purely from projection into the spin ice manifold, such that the basic result is independent of the specific model (\ref{modelice}) used to derive it. 

\subsection{Scalar Charge Theory}

For the scalar charge theory, defined by the Gauss's law $\partial_i\partial_jE^{ij} = \rho$, the low-energy Hamiltonian takes the same schematic form as in Maxwell theory, as seen in Eq.~\eqref{hamiltonian}, with the implicit gauge constraint $\partial_i\partial_jE^{ij} = 0$. As in Maxwell theory, the dispersion of the gauge modes is linear. By the same equipartition argument as before, the zero-temperature quantum correlation function satisfies
\begin{equation}
\langle E^{ij}(q)E_{ij}(-q)\rangle \propto q.
\end{equation}
To get the correct tensor structure, we start with the isotropic symmetric tensor $\frac{1}{2}(\delta^{ik}\delta^{j\ell} + \delta^{i\ell}\delta^{jk})$ and project out the $q^iq^j$ component
\begin{equation}
\label{scalarcorr}
\langle E^{ij}(q)E^{k\ell}(-q)\rangle \propto q\left(\frac{1}{2}\left(\delta^{ik}\delta^{j\ell} + \delta^{i\ell}\delta^{jk}\right) -\frac{q^iq^jq^kq^\ell}{q^4}\right),
\end{equation}
which exhibits a pinch point singularity at $q=0$, in the sense of different limiting behavior when approaching the origin from different directions. Notice, however, that the rank-4 tensor structure of the correlation function~\eqref{scalarcorr} manifests itself in a characteristic \emph{four-fold} singularity pattern, as opposed to the two-fold symmetry of pinch points in a conventional $U(1)$ spin liquid. Note also that the four-fold symmetry is present only in certain components of this rank-4 tensor e.g. $\langle E^{xx} E^{yy} \rangle$, while others such as $\langle E^{xx} E^{xx} \rangle$ only possess a two-fold symmetry. However, the presence of a four-fold symmetry in certain components of the correlator, as depicted in Fig.~\ref{fig:pinch}, will allow for easy distinction between this tensor gauge theory and more familiar spin ice models described by vector gauge theories.

In close analogy with the conventional $U(1)$ spin liquid, we can also consider a classical analogue of this tensor Coulomb phase, where only the ``spin-ice" constraint, $\partial_i\partial_jE^{ij} = 0$ is important, while the quantum splitting of degeneracies can be ignored.  By the central limit theorem, we can once again conclude that the probability distribution for $E_{ij}$ takes a Gaussian form, such that the free energy can be written as
\begin{equation}
F/T = K\int d^3x\,E^{ij}E_{ij},
\end{equation}
for some constant $K$.  The classical correlation function then takes the form
\begin{equation}
\langle E^{ij}(q)E^{k\ell}(-q)\rangle_c \propto \frac{1}{K}\left(\frac{1}{2}\left(\delta^{ik}\delta^{j\ell} + \delta^{i\ell}\delta^{jk}\right) -\frac{q^iq^jq^kq^\ell}{q^4}\right),
\end{equation}
which has the same four-fold behavior as the quantum correlation function~\eqref{scalarcorr}, but with a different, non-universal pre-factor. Again, the pinch point structure comes purely from projection into the `higher rank' spin ice manifold. 

\begin{table*}[t!]
{\renewcommand{\arraystretch}{2}%
\begin{tabular}{|c|c|c|c|}
\hline
Theory&Gauge Dispersion & Polarizations & Heat Capacity \\ \hline
Scalar Charge & $\omega \sim q$ & 5 & $C\sim T^3$ \\ \hline
Traceless Scalar Charge & $\omega \sim q$ & 4 & $C\sim T^3$ \\
\hline
Vector Charge & $\omega \sim q^2$& $3$ & $C\sim T^{3/2} $ \\ \hline
Traceless Vector Charge & $\omega \sim q^3$ & $2$ & $C\sim T$ \\\hline
\end{tabular}
}
\caption{Summary of heat capacities for the rank-2 tensor $U(1)$ spin liquids.}
\label{tab1}
\end{table*}

\subsection{Vector Charge Theory}

For the vector charge theory, defined by $\partial_i E^{ij} = \rho^j$, the low-energy Hamiltonian leads to gapless gauge modes with \emph{quadratic} dispersion $\omega\sim q^2$, unlike the previously studied theories. The implicit gauge constraint in the low-energy sector is now $\partial_i E^{ij}=0$. By the usual equipartition argument, the zero-temperature quantum correlation function satisfies
\begin{equation}
\langle E^{ij}(q)E_{ij}(-q)\rangle\propto q^2.
\end{equation}
In order to restore the tensor structure, we can start with the isotropic symmetric tensor then add terms to project off components along the $q$ direction,
\begin{align}
\begin{split}
&\langle E^{ij}(q)E^{k\ell}(-q) \rangle\propto q^2 \bigg[\frac{1}{2}(\delta^{ik}\delta^{j\ell} + \delta^{i\ell}\delta^{jk})  + \frac{q^iq^jq^kq^\ell}{q^4} \\
& - \frac{1}{2}\bigg(\delta^{ik}\frac{q^jq^\ell}{q^2} + \delta^{jk}\frac{q^iq^\ell}{q^2} + \delta^{i\ell}\frac{q^jq^k}{q^2} + \delta^{j\ell}\frac{q^iq^k}{q^2}\bigg)\bigg]. 
\end{split}
\end{align}
It can be readily checked that this expression annihilates any rank-2 tensor with a component along $q$ in either index. This correlation function once again has a pinch point singularity at $q=0$ with a characteristic four-fold symmetry, similar to that of the scalar charge theory. However, the pinch point singularity of the vector charge theory has a different power-law behavior than that of either the conventional $U(1)$ spin liquid or the scalar charge theory. The exponent with which this correlator diverges can thus be readily identified in neutron scattering data, making this type of spin liquid particularly simple to distinguish in experiments.

In close analogy with the previous section, we can also immediately write down the finite-temperature correlation function of the corresponding classical tensor Coulomb phase as
\begin{align}
\begin{split}
&\langle E^{ij}(q)E^{k\ell}(-q) \rangle\propto \frac{1}{K} \bigg[\frac{1}{2}(\delta^{ik}\delta^{j\ell} + \delta^{i\ell}\delta^{jk})  + \frac{q^iq^jq^kq^\ell}{q^4} \\
& - \frac{1}{2}\bigg(\delta^{ik}\frac{q^jq^\ell}{q^2} + \delta^{jk}\frac{q^iq^\ell}{q^2} + \delta^{i\ell}\frac{q^jq^k}{q^2} + \delta^{j\ell}\frac{q^iq^k}{q^2}\bigg)\bigg] ,
\end{split}
\end{align}
which has the same four-fold symmetry as the quantum case~\eqref{vectorcorr}.

\subsection{Traceless Theories}

For completeness, we briefly discuss the pinch point singularities of the traceless versions of the rank-2 gauge theories, which will display the same four-fold symmetry pattern.  When the scalar charge theory is given an extra tracelessness constraint $E^i_{\,\,i} = 0$, its dispersion remains linear, and the corresponding pinch points have the same scaling, $\langle E^{ij}(q)E^{k\ell}(-q)\rangle\propto q$, so this theory is not easily distinguished from its traceful cousin via neutron scattering. In contrast, when the vector charge theory has tracelessness imposed, its dispersion becomes cubic $\omega\propto q^3$. In this case, the scaling of the pinch point singularities changes to $\langle E^{ij}(q)E^{k\ell}(-q)\rangle\propto q^3$, which can be clearly distinguished in experiments from all of the previously studied $U(1)$ spin liquids.

\section{Heat Capacity}

Besides the ground state correlation functions, another useful diagnostic for certain tensor Coulomb spin liquids is the low-temperature heat capacity. Since all the emergent charges are gapped excitations, their contribution to heat capacity will be exponentially suppressed $i.e.$ follow Arrhenius behaviour, as discussed in Ref.~\cite{glassy}. Hence, the low-temperature heat capacity will be dominated by the contribution from the gapless gauge modes, which depends on both the number of gauge modes and their dispersion.

Let us assume that the gauge mode has $n_p$ independent polarizations and that its dispersion is given by $\omega \sim q^a$. Then the energy density at temperature $T$ is given by
\begin{align}
\begin{split}
E/V \sim n_p& \int d^3q\, \frac{q^a}{e^{q^a/T} -1} \\
&= \frac{4 \pi n_p}{a}\,\Gamma\left(\frac{3+a}{a}\right)\zeta\left(\frac{3+a}{a}\right)T^{\frac{3 + a}{a}},
\end{split}
\end{align}
where we have set $k_B = 1$. For the usual Maxwell theory in 3+1D, where $a=1$ and $n_p=2$, this reproduces the usual heat capacity
\beq
C_v/V = \frac{d}{dT}(E/V) = \frac{8 \pi^5 T^3}{15},
\eeq
in units where the speed of light $c=1$.

For the tensor $U(1)$ spin liquids, the results are tabulated in Table~\ref{tab1}. For both the traceful and traceless versions of the scalar charge theory $C_v \sim T^3$, which is indistinguishable from the usual Maxwell theory (and from phonon contributions) since only the numerical pre-factors are different between these, reflecting the difference in the number of independent gauge modes. In principle, one can imagine detecting the number of gauge modes in the system by deforming it along different directions by applying stress/strain, which will result in the gauge modes along that direction gaining a different dispersion. However, both the traceful and traceless versions of the vector charge theory display markedly different temperature scaling, which should serve as clear and distinctive experimental signature of these phases. 

\section{Conclusions}

In this work, we have identified several key signatures which can be used to diagnose tensor Coulomb spin liquid phases, which feature an emergent deconfined $U(1)$ symmetric tensor gauge theory. Most notably, these phases exhibit pinch-point singularities in spin-spin correlation functions, which can be easily observed in neutron scattering data. These pinch-point singularities are qualitatively different from those of a conventional $U(1)$ spin liquid, which will allow these systems to easily be distinguished from more familiar spin ice materials.  Specifically, a rank-2 tensor model has a characteristic four-fold symmetry pattern, in contrast with the two-fold symmetry of pinch points in conventional $U(1)$ spin liquids.  For tensor $U(1)$ spin liquids with rank higher than two, similar logic indicates that the pinch point singularity structure is determined by the properties of the low-energy gauge modes. For a rank $n$ theory, the resulting pinch point will have a $2n$-fold symmetry.

Additionally, we tabulated the heat capacity of various tensor Coulomb spin liquids, which provides an additional metric for diagnosing certain types of these phases. These signatures will help guide the search for material realizations of tensor Coulomb spin liquids. Apart from the diagnostics considered in this work, there remain several other features of tensor gauge theories which are expected to display behaviour distinct from that of conventional vector gauge theories. For instance, the linear response coefficients of a tensor gauge theory should provide another useful experimental metric for establishing the presence of a tensor Coulomb phase, as should the dynamical behaviour of the spin correlations (for usual vector gauge theories, this was discussed in~\cite{chalker}). These additional signatures will be discussed at length elsewhere~\cite{comingsoon}.

\section*{Acknowledgments}

We acknowledge useful conversations with T. Senthil, Mike Hermele, Han Ma, and Leo Radzihovsky. This work is partially supported by the DOE Office of Basic Energy Sciences, Division of Materials Sciences and Engineering 
under Award de-sc0010526 (S.V.); by NSF Grant 1734006 (M.P.); by a Simons Investigator Award to Leo Radzihovsky (M.P. and Y.-Z.C.); by the Foundational Questions Institute (fqxi.org; grant no. FQXi-RFP-1617) through their fund at the Silicon Valley Community Foundation (R.N.).

\end{document}